\documentclass[prb,twocolumn,10pt,aps,showpacs,groupedaddress,longbibliography]{revtex4-1}
\usepackage{graphicx}
\usepackage{amsmath}
\usepackage{amssymb}
\usepackage{bm}
\usepackage{dcolumn}
\usepackage{soul} 
\usepackage{psfrag}
\usepackage{float}
\usepackage{subeqnarray}
\usepackage{xcolor}

\usepackage{url}
\usepackage[colorlinks=true,linkcolor=blue,citecolor=blue,urlcolor=blue,breaklinks]{hyperref}

\usepackage{placeins}

\begin{document}
\author{L.-F. Zhang}
\author{V. Fern\'andez Becerra}
\author{L. Covaci}
\author{M. V. Milo\v{s}evi\'{c}}\email{milorad.milosevic@uantwerpen.be}
\affiliation{Departement Fysica, Universiteit Antwerpen,
Groenenborgerlaan 171, B-2020 Antwerpen, Belgium}
\title{Electronic properties of emergent topological defects in chiral $p$-wave superconductivity}

\begin{abstract}
Chiral $p$-wave superconductors in applied magnetic field can exhibit more complex topological defects than just conventional superconducting vortices, due to the two-component order parameter (OP) and the broken time-reversal symmetry.  We investigate the electronic properties of those exotic states, some of which contain clusters of one-component vortices in chiral components of the OP and/or exhibit skyrmionic character in the \textit{relative} OP space, all obtained as a self-consistent solution of the microscopic Bogoliubov-de Gennes equations. We reveal the link between the local density of states (LDOS) of the novel topological states and the behavior of the chiral domain wall between the OP components, enabling direct identification of those states in scanning tunneling microscopy. For example, a skyrmion always contains a closed chiral domain wall, which is found to be mapped exactly by zero-bias peaks in LDOS. Moreover, the LDOS exhibits electron-hole asymmetry, which is different from the LDOS of conventional vortex states with same vorticity. Finally, we present the magnetic field and temperature dependence of the properties of a skyrmion, indicating that this topological defect can be surprisingly large in size, and can be pinned by an artificially indented non-superconducting closed path in the sample. These features are expected to facilitate the experimental observation of skyrmionic states, thereby enabling experimental verification of chirality in emerging superconducting materials.
\end{abstract}


\maketitle
\section{Introduction}

\begin{figure*}
  \centering
  \includegraphics[width=0.9\textwidth]{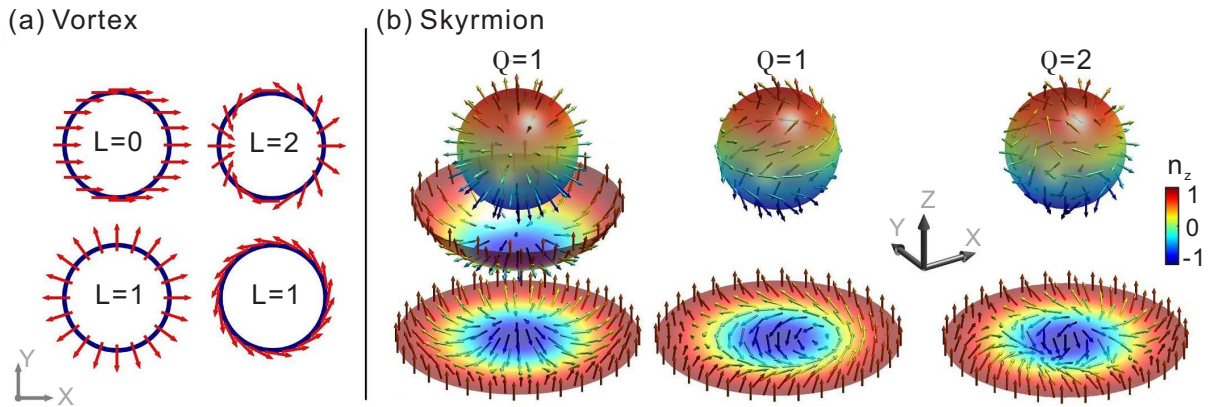}
  \caption{(Color online) Schematic order-parameter (OP) field configurations for (a) a vortex in (b) a skyrmion.  The arrows indicate the local OP field.  The color in (b) indicates the $z$-amplitude of the OP field $n_z$.  For a two-dimensional OP field space, vortex can be generated on an enclosed loop, as shown in (a), where field must turn integer number of times $L$ along the loop.  $L$ is called the winding number and it is a topological invariant.  It indicates that the wrap of the OP field can not be removed or changed to another vortex configurations with different $L$ by continually deforming without cutting the loop.  For a three-dimensional OP field space, the skyrmionic topological defect can be generated on a closed surface.  As shown in (b), the OP field must wrap integer number of times $Q$ on the surface of the sphere.  $Q$ is called the topological charge and it is also a topological invariant (similar to winding number $L$) indicating that the skyrmions with different $Q$ can not be transformed into one another by continually deforming the OP configuration.  The skyrmionic topological feature is preserved under mapping from the surface of the sphere to a plane.}
  \label{skyrmodel}
\end{figure*}
%

Topological defects play an important role in condensed matter physics, especially those which can be described and classified by homotopy groups of their order parameter (OP) space.\cite{mermin_topological_1979, braun_topological_2012}  For example, when the OP space is two-dimensional, a vortex appears as a topological defect [see Fig.~\ref{skyrmodel}(a)].  It is classified by the first homotopy group $\pi_1(S^1)\in Z$ and labeled by an integer winding number $L$.  The best known example is the Abrikosov vortex in superconductors.\cite{ketterson_superconductivity_1999}  Its main characteristics are that it has a singular vortex core and carries magnetic flux quantized in $\Phi_0=hc/2e$.  The Abrikosov vortex matter has been well studied,\cite{gygi_selfconsistent_1991, brandt_precision_1997, schweigert_vortex_1998, virtanen_multiquantum_1999, xu_magnetic_2008, zhang_unconventional_2012, zhang_vortex_2013, silaev_selfconsistent_2013} in the context of understanding detrimental effects of magnetic field on superconductivity, but also to devise various vortex-based (fluxonic) devices.\cite{golod_single_2015, milosevic_fluxonic_2007, milosevic_vortex_2010}

When the OP space is three-dimensional, a more complex topological defect, the 2D skyrmion, may arise [see Fig.~\ref{skyrmodel}(b)].  The 2D skyrmions are classified by the second homotopy group $\pi_2(S^2)\in Z$ and labeled by an integer topological charge $Q$.\cite{mermin_topological_1979, braun_topological_2012}  Such skyrmions do not exhibit singularity in the OP field.
Skyrmions are frequently observed structures in physics.  For example, the skyrmionic spin textures have been observed in magnetic systems, showing potential applications in novel spintronic devices.\cite{fert_skyrmions_2013}  Also, skyrmions have been discussed in quantum Hall systems\cite{sondhi_skyrmions_1993}, Bose-Einstein condensates\cite{leslie_creation_2009}, superfluids\cite{walmsley_chirality_2012}, and superconductors\cite{knigavko_magnetic_1999, knigavko_magnetic_1999-1, li_skyrmion_2009, li_elasticity_2007, garaud_skyrmionic_2012}, where they are formed by spin or pseudo-spin textures.

Since recently, spin-triplet chiral $p$-wave superconducting states attract great interest because of their exotic properties and the possibility to have topologically protected quantum states. \cite{volovik_universe_2009} 
Such unconventional pairing is realized in the A-phase of superfluid $^3$He and may be attributed also to the layered ruthenate superconductor Sr$_2$RuO$_4$.\cite{maeno_superconductivity_1994, lichtenberg_story_2002}
The OP of the $p$-wave pairing state is necessarily multi-component due to the extra orbital and spin degree of freedom.  In addition, the Cooper-pair with orbital angular momentum $\mathbf{L}^{orb}= 1$ breaks the time-reversal symmetry.\cite{kallin_chiral_2012, rice_sr2ruo4_1995}  These features result in rich topological defect states, of different types, with often nontrivial vorticity.  

First, there exist domain walls with spontaneous supercurrent separating domains with different degenerate time-reversal-symmetry-broken ground states.\cite{matsumoto_quasiparticle_1999}  

Second, half-quantum vortices arise due to the extra spin freedom in OP and are predicted to be thermodynamically stable in mesoscopic samples and have been detected in Sr$_2$RuO$_4$.\cite{jang_observation_2011}.\cite{volovik_universe_2009}  It is also expected that the half-quantum vortices in two-dimensional superfluids will host Majorana states at exactly zero energy as bound states inside the vortex cores.\cite{read_paired_2000}  The Majorana zero mode gives rise to non-Abelian statistics and thus can be utilized to make topological quantum computation.\cite{kitaev_faulttolerant_2003}

Third, in $p$-wave superconductivity, there exist two types of singly quantized vortices due to the broken time-reversal symmetry.\cite{matsumoto_chiral_1999} The Cooper pairs of chiral $p$-wave pairing have internal orbital angular momentum, i.e. the paired electrons are rotating.  Then, the vortex can have either the vorticity in the same direction to the angular momentum of the rotating Cooper-pair (parallel vortex), or in the opposite direction (anti-parallel vortex).  These two types of vortices have different angular momenta, causing different properties in electronic states\cite{daino_oddfrequency_2012} leading to different optical absorption,\cite{matsumoto_chiral_1999} vortex charging effect,\cite{matsumoto_vortex_2001} and surface sensitivity effect.\cite{yokoyama_chirality_2008}

Lastly, the chiral $p$-wave pairing state allows the existence of coreless vortices (CLVs) with nonzero vorticity in only one OP component\cite{sauls_vortices_2009}, which are very different from conventional singular-core vortices. The CLVs result from the extra orbital and spin degree of freedom in the OP.  In the CLVs, the $l$-vector, which points in the direction of the orbital angular momentum of the Cooper-pair, changes orientation in a continuous way, leaving a nearly homogeneous OP amplitude throughout the whole structure.  Such structures were studied before and they are referred to as Anderson-Toulouse vortices\cite{anderson_phase_1977} and Mermin-Ho vortices\cite{mermin_circulation_1976} in liquid $^3$He-A.  The CLV with doubly quantized flux has been detected in liquid $^3$He.\cite{blaauwgeers_double-quantum_2000}  In chiral $p$-wave superconductors, this doubly quantized vortex state is predicted to be energetically favorable when compared to the state with two singly quantized vortices,\cite{garaud_skyrmionic_2012, garaud_properties_2015} and should be further stabilized in the presence of mesoscopic boundaries.\cite{fernandezbecerra_vortical_2016}  The Ginzburg-Landau simulations reported the magnetic field distributions of the CLV states,\cite{garaud_skyrmionic_2012} that are still to be been observed experimentally.

Such CLVs are extremely interesting, exhibiting a variety of different aspects.  First, they are analogous to a giant vortex in $s$-wave superconductor\cite{schweigert_vortex_1998} since they contain multiple flux quanta, but exhibit a larger size.  Then, CLV is similar to a domain wall separating domains where different OP components dominate.\cite{sauls_vortices_2009}  Recently, such domain wall was found to bind half-quantum vortices, forming a structure with multiple flux quanta.\cite{garaud_skyrmionic_2012}  Finally, the $\mathbf{l}$-vector texture of a coreless vortex was characterized as a 2D skyrmion.\cite{knigavko_magnetic_1999}  The similar situation was shown in Refs.~\onlinecite{babaev_hidden_2002, garaud_skyrmionic_2012} where a pseudo-spin texture $\mathbf{n}$ of a two-component OP exhibits 2D skyrmion texture for the coreless vortex.  Although these previous studies revealed important aspects of the coreless vortices, there is still a need for a systematic study in order to enhance understanding on the coreless vortices and skyrmionic topological defects especially concerning their bound electronic states.

In this paper, we study the possible topological defect states in chiral $p$-wave superconductors, ranging from domain walls, and vortices, to coreless vortices and skyrmions, by solving the microscopic Bogoliubov-de Gennes (BdG) equations self-consistently.  The purpose of this paper is to clarify their topological properties and also to reveal their detailed electronic properties.  The bound electronic states in e.g. vortices are known to be important for many applications.\cite{caroli_bound_1964, gygi_selfconsistent_1991, virtanen_multiquantum_1999, tanaka_electronic_2002, melnikov_mesoscopic_2002, melnikov_local_2009}  For example, they determine the low-temperature behavior of the specific heat.\cite{kopnin_enhanced_2007}  In this paper, the shown results on characteristic quasiparticle excitation spectra and details of the local density of states (LDOS) of each state (especially the states associated with the skyrmion), enable their identification in e.g. scanning tunneling microscopy (STM).  Modern STM operates at spatial resolution up to $0.1~\mathrm{nm}$, and has successfully detected to date the zero bias conductance peak at the vortex core,\cite{suderow_imaging_2014} phase transition between multi- and giant vortex states,\cite{cren_ultimate_2009, cren_vortex_2011} proximity effect\cite{serrier-garcia_scanning_2013}, Josephson vortices,\cite{roditchev_direct_2015, yoshizawa_imaging_2014} etc.  Hence our results will provide valuable info for direct detection of novel topological states, which can in turn serve as a `smoking gun' for $p$-wave superconductivity in the studied system.

The paper is organized as follows.  
In Sec.~\ref{sec:2.1} we introduce our theoretical methodology for chiral $p$-wave superconductors.  
In Sec.~\ref{sec:2.2} we define the skyrmionic topological defects in the relative OP space in two-component OP systems.  
In Sec.~\ref{sec:3} we summon the results for three distinct states without a skyrmionic topology.  Those are the vortex-free state, the parallel vortex state and the anti-parallel vortex state. 
In Sec.~\ref{sec:4} we present results on coreless vortex states.  Their OP structures, supercurrent distribution, energy spectra and LDOS are discussed.  We show that they are associated with skyrmionic topological defects in relative OP space.
In Sec.~\ref{sec:5} we reveal the magnetic field and temperature dependence of the properties of the skyrmion, followed by the investigation of an effective skyrmion pinning in Sec.~\ref{sec:6}.  
Finally, our findings are summarized in Sec.~\ref{sec:7}.

\section{Theoretical Formalism}\label{sec:2}

\subsection{Bogoliubov-de Gennes equations for chiral $p$-wave superconductors}\label{sec:2.1}

We consider chiral $p$-wave superconductors whose order parameter (OP) is expressed as
\begin{equation}\label{Eq:OPvpm}
	\mathbf{\Delta}(\mathbf{r},\mathbf{k})=\Delta_+(\mathbf{r})Y_+(\mathbf{k})+\Delta_-(\mathbf{r})Y_-(\mathbf{k}).
\end{equation}
Here the $\Delta_{\pm}(\mathbf{r})$ are the real spatial $p_x \pm ip_y$-wave OP and $Y_{\pm}(\mathbf{k})=(k_x \pm ik_y)/k_F$ are the pairing functions in relative momentum space.  We consider a disk geometry with radius $R$.  The corresponding $p_x \pm ip_y$-wave BdG equations are written as: \cite{matsumoto_vortex_2001}
\begin{equation}\label{Eq:BdG}
	\begin{bmatrix}
		H_e(\mathbf{r}) & \Pi(\mathbf{r}) \\
		-\Pi^*(\mathbf{r}) & -H_e^*(\mathbf{r})
	\end{bmatrix}
	\begin{bmatrix}
		u_n(\mathbf{r}) \\
		v_n(\mathbf{r})
	\end{bmatrix}
	= E_n
	\begin{bmatrix}
		u_n(\mathbf{r}) \\
		v_n(\mathbf{r})
	\end{bmatrix},
\end{equation}
where
\begin{equation}\label{Eq:He}
	H_e(\mathbf{r})= \frac{1}{2m}[\frac{\hbar}{i}\nabla-\frac{e}{c}\mathbf{A}(\mathbf{r})]^2-E_F
\end{equation}
is the single particle Hamiltonian with $m$ being the electron mass, $E_F$ the Fermi energy and $\mathbf{A}(\mathbf{r})$  the vector potential (we use the
gauge $\nabla \cdot \mathbf{A} = 0$).  For simplicity, we take the cylindrical two dimensional Fermi surface.  The term $\Pi(\mathbf{r})$ is written as
\begin{equation}\label{Eq:BigPi}
	\Pi(\mathbf{r}) = -\frac{i}{k_F} \sum_{\pm} [\Delta_{\pm}\square_{\pm} + \frac{1}{2}(\square_{\pm}\Delta_{\pm})],
\end{equation}
with $\square_{\pm}=e^{\pm i \theta} (\partial_r \pm \frac{i}{r}\partial_\theta )$ in cylindrical coordinates. $u_n(\mathbf{r})$($v_n(\mathbf{r})$) are electron(hole)-like quasi-particle eigen wavefunctions with the normalization condition
\begin{equation}\label{Eq:normuv}
	\int |u_n(\mathbf{r})|^2+|v_n(\mathbf{r})|^2 d\mathbf{r}=1,
\end{equation}
and $E_n$ are the corresponding quasiparticle eigenenergies.  The boundary conditions for the wavefunctions are $u_n(r=R)=0$ and $v_n(r=R)=0$.  The $\Delta_{\pm}(\mathbf{r})$ satisfy the self-consistent gap equations
\begin{equation}\label{Eq:DxDy}
	\begin{split}
		\Delta_{\pm}(\mathbf{r}) = &-i\frac{g}{2k_F}\sum_{E_n<\hbar \omega_D} [v_n^*(\mathbf{r})\square_{\mp}u_n(\mathbf{r})- \\ & u_n(\mathbf{r})\square_{\mp}v_n^*(\mathbf{r})]\times [1-2f(E_n)],
	\end{split}
\end{equation}
where $k_F=\sqrt{2mE_F/\hbar^2}$ is the Fermi wave length, $g$ the coupling constant and $f(E_n)=[1+\exp(E_n/k_B T)]^{-1}$ is the Fermi distribution function.  The summations in Eq.~(\ref{Eq:DxDy}) are over all the quasiparticle states with energies in the Debye window $\hbar \omega_D$.  The supercurrent density is calculated by
\begin{equation}\label{Eq:SuperC}
	\begin{split}
		\mathbf{j}(\mathbf{r}) =& \frac{e\hbar}{2mi} \sum_n \bigg \{ f_n u_n^*(\mathbf{r}) \Big [ \nabla- \frac{ie}{\hbar c} \mathbf{A}(\mathbf{r}) \Big ] u_n(\mathbf{r}) \\
		& + (1-f_n) v_n(\mathbf{r}) \Big [ \nabla- \frac{ie}{\hbar c} \mathbf{A}(\mathbf{r}) \Big ] v_n^*(\mathbf{r}) - h.c. \bigg \}.
	\end{split}
\end{equation}

In order to perform the self-consistent simulation, we include the contribution of the supercurrent to the total magnetic field.  Then, the vector potential $\mathbf{A}(\mathbf{r})$ in Eqs.~(\ref{Eq:He}) and (\ref{Eq:SuperC}) has two parts, i.e. $\mathbf{A}(\mathbf{r}) = \mathbf{A_0}(\mathbf{r}) + \mathbf{A_1}(\mathbf{r})$, where $\mathbf{A_0}(\mathbf{r}) = \frac{1}{2} H_0r \mathbf{e}_\theta$ corresponds to the applied magnetic field $\mathbf{H}=H_0\mathbf{e}_z$ and the $\mathbf{A_1}(\mathbf{r})$ is induced by the supercurrent and obey the Maxwell equation
\begin{equation}\label{Eq:JtoA}
	\nabla \times \nabla \times \mathbf{A_1}(\mathbf{r}) =\frac{4\pi}{c} \mathbf{j}(\mathbf{r}).
\end{equation}
However, we find that the $\mathbf{A_1}(\mathbf{r})$ is negligible due to the very thin superconducting sample.  As a result, the contribution of the supercurrent to the total magnetic field can be completely neglected in this type of simulation.

In this paper, we only consider vortex and skyrmion states with cylindrical symmetry.  Therefore, the $p_x \pm ip_y$ components of the order parameter are expressed as $\Delta_{\pm}(\mathbf{r}) = \Delta_{\pm}(r) e^{iL_{\pm}\theta}$ with winding numbers $L_{\pm}$, respectively.  Due to operators $\square_{\pm}$ in Eqs.~(\ref{Eq:BdG})-(\ref{Eq:DxDy}), $\Delta_{\pm}$ have a $\pm 1$ Cooper-pair phase winding, respectively, leading to $L_-=L_++2$.  This also breaks the time-reversal symmetry, resulting in chiral states.

In a cylindrical system, the quasiparticle wavefunctions $u_n(\mathbf{r})$ and $v_n(\mathbf{r})$ can be expanded in terms of the following Bessel set\cite{gygi_selfconsistent_1991}:
\begin{equation}\label{Eq:Bessel1}
	\begin{pmatrix}
		u_n(\mathbf{r})\\
		v_n(\mathbf{r})
	\end{pmatrix}
	=
	\sum_j
	\begin{pmatrix}
		c_{n\mu j}\varphi_{j\mu}(r)e^{i\mu\theta} \\
		d_{n\mu' j}\varphi_{j\mu'}(r)e^{i\mu'\theta}
	\end{pmatrix},
\end{equation}
where $c_{n\mu j}$ and $d_{n\mu' j}$ are coefficients, $\mu$,$\mu' \in \mathbb{Z}$ are angular quantum numbers corresponding to the angular momentum, and
\begin{equation}\label{Eq:Bessel2}
	\varphi_{j\mu}(r)=\frac{\sqrt{2}}{RJ_{\mu+1}(\alpha_{j\mu})}J_{\mu}(\alpha_{j\mu}\frac{r}{R}),
\end{equation}
with $J_{\mu}$ the $\mu$th Bessel function and $\alpha_{j\mu}$ the $j$th zero of $J_{\mu}$.  Note that $\mu'=\mu-L_+-1$  because of the phase winding in $\Delta_{\pm}$, i.e. $L_-=L_++2$.  Then, the BdG equations are reduced to a matrix eigenvalue problem and can be solved separately in each subspace of fixed $\mu$ and $\mu'$.

After the self-consistent solutions are obtained, we calculate the LDOS as usual
\begin{equation}\label{Eq:LDOS}
	A(\vec{r},E)=\sum_n[|u_n(\vec{r})|^2\delta(E-E_n)+|v_n(\vec{r})|^2\delta(E+E_n)].
\end{equation}

For each quasiparticle state, we can define the spectral weight $Z_n$:
\begin{equation}\label{Eq:Zn}
	Z_n=\int|u_n(\mathbf{r})|^2 d\mathbf{r}.
\end{equation}
$Z_n \in [0,1]$ and it represents the contribution of the electronic part of the wave function of a Bogoliubov quasiparticle state. A state with $Z_n<0.5$ indicates a hole-like state while $Z_n>0.5$ is an electron-like state.   A Bogoliubov quasiparticle state is well formed when it couples between half-electron and half-hole, i.e. for $Z_n =0.5$.

Next, we remark that the quasi-particle states have the following time-reversal relation:
\begin{equation}\label{Eq:BdG_TR1}
	\{ u_{-E_n},v_{-E_n} \} = \{ v^*_{E_n},u^*_{E_n} \}.
\end{equation}
It indicates that a state having energy $E_n$ and angular momentum $(\mu,\mu')$ carries the same information as a state having energy $-E_n$ and angular momentum $(-\mu',-\mu)$. This allows us to reduce half of the computational time by only considering half of the angular momentum $(\mu,\mu')$.  Due to this, it is sufficient to display the quasiparticle excitation spectrum with both positive and negative energy $E_n$ but with only positive angular momentum $\mu$ or $\mu'$.

We also remark that our chiral $p$-wave BdG equations are invariant under the time-reversal operations:
\begin{equation}\label{Eq:BdG_TR2}
	\{ \Delta_{\pm}, \mathbf{B} \} \rightarrow \{ \Delta^*_{\mp}, -\mathbf{B} \},
\end{equation}
where $\mathbf{B}$ is the magnetic field.  In the bulk the two degenerate ground states are the $p_x+ip_y$ and $p_x-ip_y$-wave states.  At zero temperature, their OP $( \Delta_+, \Delta_- )= \Delta_0 (1,0)$ and $\Delta_0 (0,1)$, respectively, where $\Delta_0 \in \mathbb{R}$ is the bulk OP at zero temperature.  These two states can be mirrored by Eq.~(\ref{Eq:BdG_TR2}).  The situation is the same for vortex states.  For example, when one knows the $\Delta_+$ dominant vortex states with winding numbers $( L_+, L_- )$, one can easily obtain the $\Delta_-$ dominant vortex states with winding numbers $( -L_-, -L_+ )$ by using Eq.~(\ref{Eq:BdG_TR2}).  The complete study requires to consider both $\Delta_+$ dominant and $\Delta_-$ dominant states for all possible (positive and negative) winding numbers.  However,  with the time-reversal operations of Eq.~(\ref{Eq:BdG_TR2}), it is equivalent to consider only half of the possible winding numbers but for both $\Delta_+$ dominant and $\Delta_-$ dominant states.

Next we define the $p_x$ and $p_y$-wave OP $\Delta_x$ and $\Delta_y$.  They often show interesting properties and can provide important information about the vortex and skyrmion states.  The OP expressed by $\Delta_x$ and $\Delta_y$ can be written as
\begin{equation}\label{Eq:OPvxy1}
	\mathbf{\Delta}=(\Delta_x k_x+\Delta_y k_y)/k_F.
\end{equation}
Eq.~(\ref{Eq:OPvpm}) can also be expressed as
\begin{equation}\label{Eq:OPvxy2}
	\mathbf{\Delta}=\{[\Delta_++\Delta_-]k_x+i[\Delta_+-\Delta_-]k_y \}/k_F.
\end{equation}
By comparing Eqs.~(\ref{Eq:OPvxy1}) and (\ref{Eq:OPvxy2}), we find 
\begin{equation}\label{Eq:Dmp2Dxy}
	\begin{split}
		\Delta_x &= \Delta_+ + \Delta_-,\\
		\Delta_y &= i(\Delta_+ - \Delta_-).
	\end{split}
\end{equation}
%

\subsection{Skyrmionic character in relative order parameter space}\label{sec:2.2}

In a two-component OP system, a 2D skyrmionic texture is not obvious by looking at the OP configurations.  However, it can be well-understood by projecting the system onto a pseudo-spin space.  In this section, we show that the pseudo-spin space is actually a \textit{relative OP space} where both relative amplitude and relative phase between the two OP components play an important role.  The relative OP space leads to skyrmion, which does not occur in a one-component OP system.  As a result, both winding numbers of each OP component and the topological charge associated with the skyrmion are necessary to describe the superconducting state.

We start from a complex two-component OP field $\mathbf{\Delta}(\mathbf{r}) = (\Delta_1(\mathbf{r}), \Delta_2(\mathbf{r}))^T = (|\Delta_1|e^{i\theta_1}, |\Delta_2|e^{i\theta_2})^T$ where $|\Delta_i|$ and $\theta_i$ are the amplitude and phase of the $\Delta_i$ component, respectively.  We decompose the OP as $\mathbf{\Delta}(\mathbf{r})= |\mathbf{\Delta}(\mathbf{r})| \boldsymbol{\mathbf{\chi}}(\mathbf{r})$ where $|{\Delta} (\mathbf{r})| =\sqrt{|\Delta_1|^2+|\Delta_2|^2}$ is the total OP amplitude and $\boldsymbol{\mathbf{\chi}} (\mathbf{r})=(|\chi_1|e^{i\theta_1}, |\chi_2|e^{i\theta_2})^T$ is the normalized complex-valued spinor satisfying $|\chi_1|^2+|\chi_2|^2=1$.  The $\boldsymbol{\mathbf{\chi}}(\mathbf{r})$ generates a four-dimensional OP space and it points to the surface of a solid unit sphere in four-dimensional space $S^3$.

Next we compare the $\boldsymbol{\mathbf{\chi}} (\mathbf{r})$ with pesudo-spin $\mathbf{n}$.  Following Ref.~\onlinecite{babaev_hidden_2002}, the pseudo-spin $\mathbf{n}$ is defined as: 
\begin{equation}\label{Eq:pesudospin}
	\mathbf{n} = (n_x, n_y, n_z) = \frac{\mathbf{\Delta}^\dagger \vec{\sigma} \mathbf{\Delta}}{\mathbf{\Delta}^\dagger \mathbf{\Delta}} = \boldsymbol{\mathbf{\chi}}^\dagger \vec{\sigma} \boldsymbol{\mathbf{\chi}},
\end{equation}
where $\vec{\sigma}$ is the Pauli matrices and $\mathbf{n}$ is a 3D unit vector $|\mathbf{n}|=1$, and points to the surface of a solid unit sphere in three-dimensional space $S^2$.

It is worth noting that $\mathbf{n}$ is a gauge invariant field.  Let $\boldsymbol{\mathbf{\widetilde{\chi}}}= e^{i\theta} \boldsymbol{\mathbf{\chi}}$, the $\mathbf{n}$ does not change, i.e. 
\begin{equation}\label{Eq:pesudospin-2}
	\mathbf{n} = \mathbf{\widetilde{\chi}}^\dagger \vec{\sigma} \mathbf{\widetilde{\chi}} =\mathbf{\chi}^\dagger \vec{\sigma} \mathbf{\chi}.
\end{equation}
We take $\theta=-\theta_1$ so that $\boldsymbol{\mathbf{\widetilde{\chi}}}$ can be reduced to a three-dimensional field, i.e. 
\begin{equation}\label{Eq:pesudospin-3}
	\boldsymbol{\mathbf{\widetilde{\chi}}} =(|\chi_1|, |\chi_2|e^{i\phi}) = (\cos\alpha, \sin\alpha \cos \phi, \sin\alpha \sin\phi),
\end{equation}
where $\phi = \theta_2-\theta_1$ is the \textit{relative phase} and $\alpha = \tan^{-1} |\chi_2|/|\chi_1|$ represents the \textit{relative amplitude}.  Note that $0 \leqslant \alpha \leqslant \pi/2$ indicates that  $\boldsymbol{\mathbf{\widetilde{\chi}}}$ is a set of points on a unit hemisphere.
In the last step, we rotate $\boldsymbol{\mathbf{\widetilde{\chi}}}$ globally by an angle $-90^{\circ}$ about the $y$-axis in order to match the orientation of pseudo-spin vector $\mathbf{n}$.  Finally, we reach 
\begin{equation}\label{Eq:pesudospin-4}
	\boldsymbol{\mathbf{\widetilde{\chi}}} = (\sin\alpha \cos \phi, \sin\alpha \sin\phi, \cos\alpha).
\end{equation}

A straightforward calculation of the pseudo-spin vector $\mathbf{n}$ results in 
\begin{equation}\label{Eq:pesudospin-5}
	\mathbf{n} = (\sin2\alpha \cos \phi, \sin2\alpha \sin\phi, \cos2\alpha),
\end{equation}
with $0 \leqslant 2\alpha \leqslant \pi$.  It is clear that $\mathbf{n}$ has the same structure as $\boldsymbol{\mathbf{\widetilde{\chi}}}$, except for the azimuthal angle, $\alpha \rightarrow 2\alpha$, so that $\mathbf{n}$ is a set of points of a whole unit sphere $S^2$.  As a result, the pseudo-spin vector $\mathbf{n}$ represents the \textit{relative OP space} as well.  

Such a three-dimensional real vector field $\mathbf{n}$ can exhibit non-trivial skyrmionic topological defects on an enclosed surface such as $S^2$, as shown e.g. in Fig.~\ref{skyrmodel}(b).  Just like the winding number (vorticity) for vortex matter, the skyrmion is described by its topological charge $Q$, which is calculated as
\begin{equation}\label{Eq:TopoCharge}
	Q=\int \mathbf{n} \cdot (\partial_x \mathbf{n} \times \partial_y \mathbf{n}) ~ dxdy,
\end{equation}
and counts the number of times that $\mathbf{n}$ wraps the enclosed surface.  It is classified by the second homotopy group $\pi_2(S^2)\in Z$, where $Z$ is an integer.

It is worth to mention that the two-component OP field $\mathbf{\Delta}(\mathbf{r})$ can be separated into three parts:  $\Delta$ the total amplitude, $e^{i\theta}$ the common phase term, and $\mathbf{n}$ representing the relative OP space.  The total amplitude $\Delta$ does not contain any topological structure due to its scalar nature.  The common phase term $e^{i\theta}$ has two effects: 1) the global $U(1)$ gauge invariance for superconductivity; 2) the common phase distributions and the common winding number $L$.  Actually, the first two terms represent a one-component OP system.  The third part $\mathbf{n}$ arises an effect resulting from the two-component OP.  It is different from the phase soliton where only the relative phase is taken into account.\cite{tanaka_soliton_2001, lin_phase_2012}  The $\mathbf{n}$ induces extra skyrmionic topological defects, labelled by the topological charge $Q$.  

Finally, one sees that the two-component OP system contains additional topological possibilities next to just vortices, where winding numbers $(L_1,L_2)$ of both OP components and the topological charge $Q$ are all necessary to describe the superconducting state.  Therefore in this manuscript, we use $(L_1,L_2,Q)$ to label different observed states.

\section{Structure of vortex states without skyrmionic topology}\label{sec:3}

In this section, we investigate three prominent vortex states not exhibiting a skyrmionic topology: Vortex-free state $(L_+, L_-,Q)=(0,2,0)$, parallel vortex state $(1,3,0)$ and anti-parallel vortex state $(-1,1,0)$.  Since $Q=0$ for all these states, we omit it in this section.  The OP structures, supercurrent density, quasiparticle excitation spectrum $E_n$, and LDOS for the considered states will be presented, where some findings coincide with previous works.\cite{matsumoto_quasiparticle_1999, sauls_surface_2011}  In our analysis, we found that the $p_x$ and $p_y$ OP components $\Delta_x$ and $\Delta_y$ are very useful, and will be employed in the analysis of the found vortex states.  

The calculations are performed for the sample of radius $R = 51\xi_0$, where $\xi_0=\hbar v_F/\pi\Delta_0$ is the BCS coherence length at zero temperature, with $v_F$ the Fermi velocity and $\Delta_0$ the bulk OP at zero temperature.  $E_F=\hbar \omega_D$ and $\hbar \omega_D/\Delta_0 \approx 14$, resulting in $k_F\xi_0 = 9$.  We also set the applied magnetic field to $\mathbf{H}=0$, so  the reported properties are surly not a consequence of the magnetic field.  The considered temperature is $T=0.1T_c$.  The results remain qualitatively the same when we change the magnetic field $\mathbf{H}$ and temperature $T$.

\begin{figure}
	\centering
	\includegraphics[width=\columnwidth]{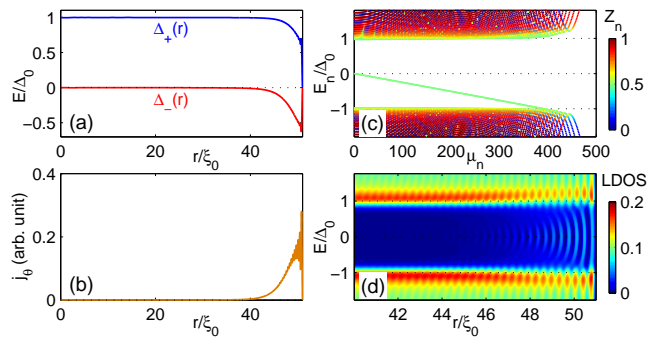}
	\caption{(Color online) Vortex-free state $(L_+, L_-)=(0,2)$ with $\Delta_+$ dominant.  (a) Profile of $\Delta_{\pm}(r)$ at $\theta=0$. (b) Azimuthal supercurrent density $j_{\theta}(r)$. (c) The quasiparticle excitation spectrum $E_n$ as a function of the positive angular momentum $\mu$.  The negative part of the spectrum can be obtained by the time-reversal relation of Eq.~(\ref{Eq:BdG_TR1}).  The color coding indicates the spectral weight $Z_n$. (d) The LDOS near surface as a function of radius $r$ and bias energy $E$.}
	\label{v0main}
\end{figure}
\begin{figure*}
	\centering
	\includegraphics[width=\textwidth]{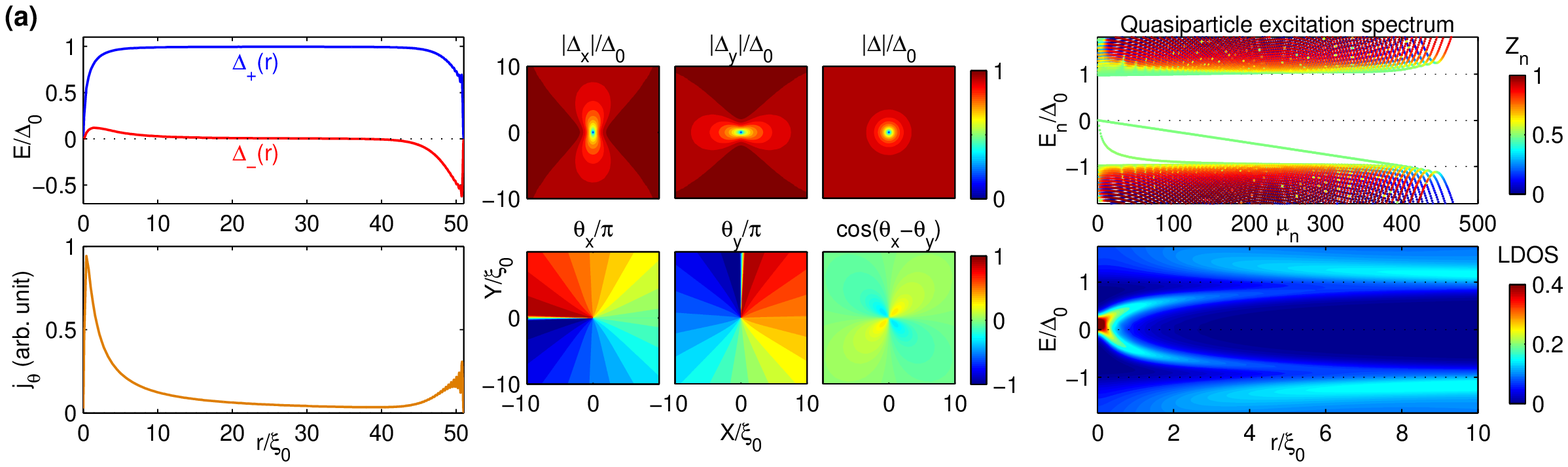}
	\includegraphics[width=\textwidth]{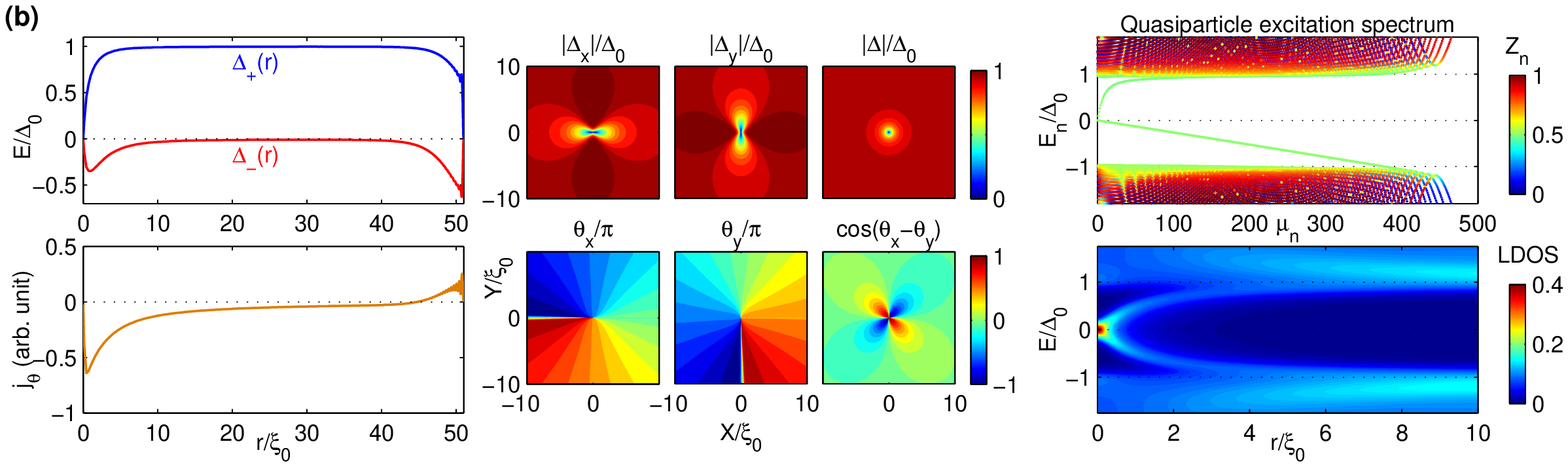}
	\caption{(Color online) Two types of $\Delta_+$-dominant singly-quantized vortex states: $(L_+, L_-)=(1,3)$ and $(-1,1)$, respectively shown in panels (a) and (b).  Plots on the left show profiles of $\Delta_{\pm}(r)$ and the azimuthal supercurrent density $j_{\theta}(r)$. Central plots show both amplitude and phase of OP components $\Delta_x(\mathbf{r})$ and $\Delta_y(\mathbf{r})$, their relative phase $\cos(\theta_x-\theta_y)$, and the total OP amplitude $|\Delta(\mathbf{r})|$.  Note that the winding numbers of $\Delta_x$ and $\Delta_y$ are $L_x=L_y=1$ for the $(L_+, L_-)=(1,3)$ state and $L_x=L_y=-1$ for the $(L_+, L_-)=(-1,1)$ state.  Plots on the right show the quasiparticle excitation spectrum $E_n$ as a function of the angular momentum $\mu$ (with color coding indicating the spectral weight $Z_n$), and the LDOS around the vortex core as a function of radial distance $r$ and bias energy $E$.}
	\label{singlemain}
\end{figure*}
%

We first introduce the vortex-free state $(L_+, L_-)=(0,2)$, with $\Delta_+$ as a dominant component.  The results are summarized in Fig.~\ref{v0main}.  The state is analogous to the Meissner state in $s$-wave superconductors, therefore it is the first step for understanding vortex and skyrmion states.  In bulk, the ground state is $(\Delta_+, \Delta_-)=\Delta_0(1,0)$.  However, the physical properties significantly change near a surface.\cite{matsumoto_quasiparticle_1999}  As seen from Fig.~\ref{v0main}(a), the $|\Delta_+|$ suppresses and $|\Delta_-|$ rises at the surface, where an anticlockwise supercurrent is also induced [see Fig.~\ref{v0main}(b)].  The quasiparticle excitation spectrum shown in Fig.~\ref{v0main}(c) reveals chiral surface states with a linear dispersion around the Fermi energy.\cite{matsumoto_quasiparticle_1999, stone_edge_2004, sauls_surface_2011}  These are Andreev bound states induced by the chirality of the superconducting state.\cite{furusaki_spontaneous_2001}  The states cross the Fermi energy but there is no exact-zero energy Majorana mode.\cite{stone_edge_2004}  They contribute to the low-bias LDOS distributions near the surface, as shown in Fig.~\ref{v0main}(d). Note that the LDOS and the supercurrent $j_{\theta}(r)$ show Friedel-like oscillations with a wave vector $2k_F$ near the surface.

Here we note that the spontaneous surface supercurrent is the major characteristic of the superconducting state with broken time-reversal symmetry.  Experiments to date have observed the surface bound states\cite{kashiwaya_edge_2011} but failed to capture the surface supercurrent.\cite{kirtley_upper_2007, hicks_limits_2010, curran_vortex_2011}  One possible explanation is that the supercurrent depends on exact geometry and band structure of the sample,\cite{bouhon_current_2014} but that discussion is out of the scope of this paper.


Next we present the case of two types of singly quantized vortex states with $\Delta_+$ dominant:  the parallel vortex state $(L_+, L_-)=(1,3)$ and the anti-parallel vortex state $(L_+, L_-)=(-1,1)$, shown in Fig.~\ref{singlemain}(a) and (b), respectively.  Here we remind the reader that the vortex and the anti-vortex states exhibit very different properties due to the broken time-reversal symmetry.\cite{matsumoto_chiral_1999, matsumoto_vortex_2001, sauls_surface_2011}

The left plots in Fig.~\ref{singlemain}(a,b) show $\Delta_{\pm}(r)$ and the supercurrent density profile $j_{\theta}(r)$.  Compared to the vortex-free $(L_+, L_-)=(0,2)$ state shown in Fig.~\ref{v0main}, $\Delta_{+}(r)$ exhibits a singular vortex core in the center of the sample.  At the same time, $\Delta_{-}(r)$ is induced near the vortex core and also exhibits singularity there, so the cores in $\Delta_{\pm}$ overlap.  However, the two possible singly-quantized vortex states have different vortex core structures.  For the parallel vortex $(1,3)$ state, $\Delta_{\pm}(r)$ show different asymptotic behavior: $\Delta_+(r) \propto r$ while $\Delta_-(r) \propto r^3$.  For the anti-vortex $(-1,1)$ state, both $|\Delta_{\pm}(r)| \propto r$.  In addition, the states have different supercurrent density distributions.  The parallel vortex $(1,3)$ state has the positive vorticity, leading to the clockwise $j_{\theta}(r)$ around the vortex.  In contrast, the anti-vortex $(-1,1)$ state has the negative vorticity, leading to the anti-clockwise $j_{\theta}(r)$ around the vortex core.

Previous works concerning vortex states in chiral $p$-wave superconductors rarely presented the $p_x$ and $p_y$ OP components $\Delta_x$ and $\Delta_y$.  We actually found that they can be very useful in the analysis of interesting properties, especially related to the vorticity of the sample.  The central plots in Fig.~\ref{singlemain} show the profiles of $\Delta_x$, $\Delta_y$,  the relative phase between them $\theta_x-\theta_y$, and the total OP $\Delta$.  We find the winding numbers $L_x=L_y=1$ for the parallel vortex $(1,3)$ state and $L_x=L_y=-1$ for the anti-vortex $(-1,1)$ state, thus better describing the vorticity of the sample than the angular momenta of $\Delta_\pm$.  The vortex cores in $\Delta_x$ and $\Delta_y$ are at the sample center and they overlap.  Unlike the cylindrical vortex core structures in $\Delta_{\pm}$, the vortex cores are deformed in $\Delta_x$ and $\Delta_y$, and exhibit different profiles for the $(1,3)$ and $(-1,1)$ states.  It is interesting that $\Delta_y$ can be obtained by rotating $\Delta_x$ with $90$ degrees clockwise for the $(1,3)$ state and anticlockwise for the $(-1,1)$ vortex state.  It is also interesting to note that the relative phase $\theta_x-\theta_y$ twirls twice for both cases, exhibiting a cloverleaf profile.  For the $(-1,1)$ vortex state, $\Delta_x$ and $\Delta_y$ alternate between being fully in-phase and fully out-of-phase around the vortex core.

The right hand side plots in Fig.~\ref{singlemain} show the quasiparticle excitation spectrum $E_n(\mu_n)$ and the LDOS.  Comparing to the vortex-free $(L_+, L_-)=(0,2)$ state, one more branch of bound states appears within the gap energy $\Delta_0$ in the excitation spectrum.  Those are the vortex bound states, localized around the vortex core.\cite{matsumoto_chiral_1999}  
The vortex bound states for the $(1,3)$ and $(-1,1)$ states are different.  For $(1,3)$ vortex states, the bound states reside in the negative energy range for positive angular momentum $\mu_n$.  However, for the $(-1,1)$ state they have positive energy for positive $\mu_n$, due to opposite vorticity.

It was demonstrated in Ref.~\onlinecite{tewari_index_2007, gurarie_zero_2007, mizushima_vortex_2010} that there exists a pair of zero-energy Majorana modes for a single vortex with odd vorticity in the chiral $p$-wave superconductivity.  The energy levels of the vortex bound states appear at integer points $E_n \sim n E_{\delta}$, where $n$ is an integer and $E_\delta$ is the level spacing of the order of $\Delta^2_0/E_F$.\cite{takigawa_vortex_2001}  For the state with $E_n=0$, the time-reversal relation of Eq.~(\ref{Eq:BdG_TR1}) prescribes the zero-energy state appearing as a pair, and the quasiparticle wave functions keep the relation $u_n(\mathbf{r})=v_n^*(\mathbf{r})$.  Thus, the quasiparticle creation operator is equivalent to the annihilation of a quasiparticle, which corresponds to the Majorana fermions.\cite{gurarie_zero_2007}  However, the Majorana zero mode splits when there exists vortex-vortex interaction or/and vortex-surface interaction.\cite{mizushima_splitting_2010}  In our case where $R=51\xi_0$, the energies of the lowest vortex bound state of both cases are of the order of $10^{-7}\Delta_0$.  It indicates the existence of the Majorana zero mode and the vortex-surface interaction being negligible.  With sample radius $R$ decreasing, the energy of the lowest vortex bound state oscillates and its envelope increases with exponential law.
The vortex bound states of both cases are the well-formed Bogoliubov quasiparticle states with $Z_n=0.5$, which is also supporting the Majorana zero mode.

The LDOS showing in Fig.~\ref{singlemain} reveals the zero-bias peak at the vortex core, corresponding to the same characteristic of vortex states with odd winding number in $s$-wave superconductors.  It is worth noting that the LDOS is asymmetric for $E \leftrightarrow -E$ for the $(1,3)$ state and symmetric for the $(-1,1)$ state.

\section{Structure of skyrmionic topological defects}\label{sec:4}

\begin{figure*}
	\centering
	\includegraphics[width=0.9\linewidth]{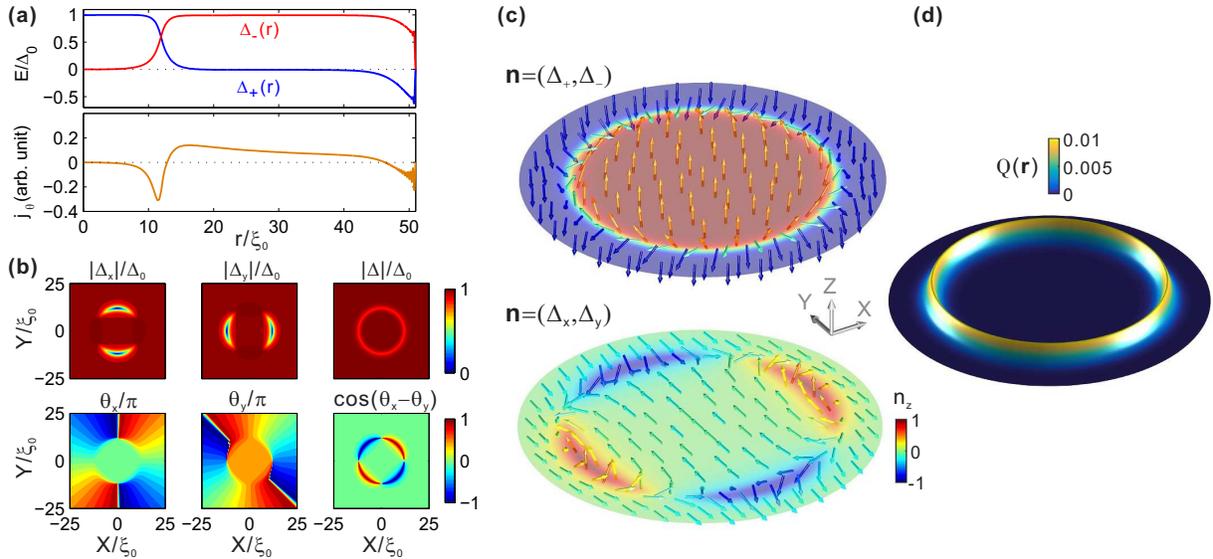}
	\caption{(Color online) Topological structure of the skyrmion state $(L_+, L_-, Q)=(0,2,2)$.  (a) Profiles of $\Delta_{\pm}(r)$ and the azimuthal supercurrent density $j_{\theta}(r)$. (b) The amplitude and the phase of $\Delta_x(\mathbf{r})$ and $\Delta_y(\mathbf{r})$, their relative phase $\cos(\theta_x-\theta_y)$, and the total OP amplitude $|\Delta(\mathbf{r})|$. Note that the winding numbers of $\Delta_x$ and $\Delta_y$ are $L_x=L_y=2$.  (c) The texture $\mathbf{n}(\mathbf{r})$ of the relative OP space, calculated using $\Delta_{\pm}$ (upper panel), and using $\Delta_{x}$ and $\Delta_{y}$ (lower panel).  The colors show the amplitude of the $z$-component of $\mathbf{n}(\mathbf{r})$.  Both shown pseudospin textures give topological charge density $Q(\mathbf{r})$ shown in panel (d) and the topological charge $Q=2$.}
	\label{skyr_tp}
\end{figure*}
\begin{figure*}
	\centering
	\includegraphics[width=0.9\textwidth]{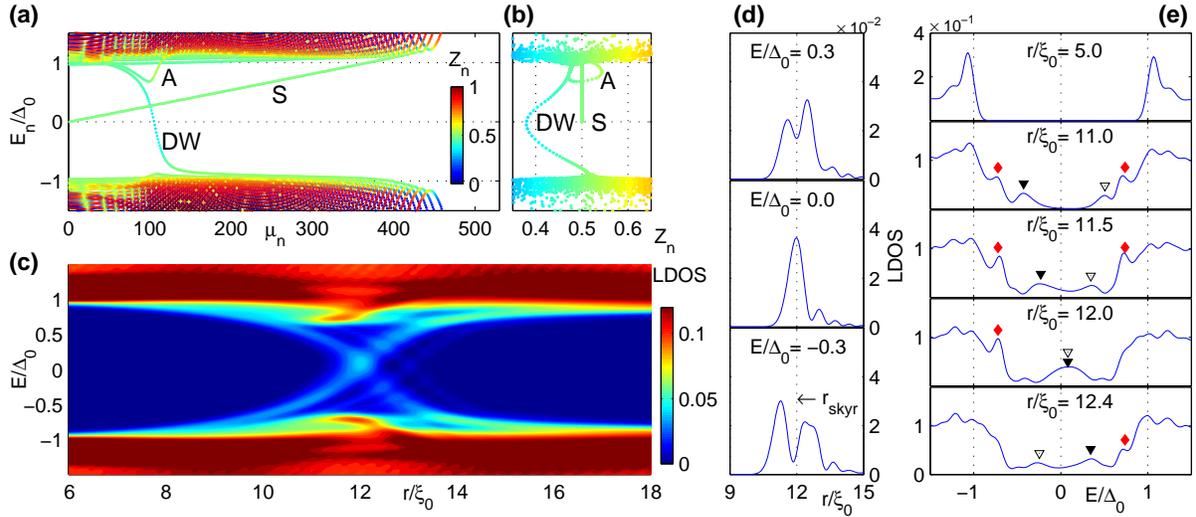}
	\caption{Electronic structure of the skyrmion state $(L_+, L_-, Q)=(0,2,2)$.  (a) The quasiparticle excitation spectrum $E_n$ as a function of the angular momentum $\mu_n$ (color coding indicates the spectral weight $Z_n$).  S, DW and A represent the surface bound state, domain-wall bound state and the Andreev bound state associated with the domain wall, respectively.  Their spectral weights $Z_n$ are shown in panel (b).  (c) The LDOS$(r,E)$ around the skyrmion as a function of radial distance $r$ and bias energy $E$.  (d) The profiles of the LDOS$(r)$ around the domain wall at bias energies $E=0.3$, $0$ and $-0.3$.  The chiral domain wall is at $r_{skyr}/\xi_0=12$.  (e) The profiles of the LDOS$(E)$ as a function of bias energies $E$ at several radial distances.  The peaks labeled by triangles (diamonds) are induced by the domain-wall bound states (Andreev bound states). }
	\label{skyr_elek}
\end{figure*}

Coreless vortices are one of the most striking states emerging in the chiral $p$-wave superconductivity.  They exhibit an additional topology which is skyrmionic.  The one known coreless vortex state is the doubly quantized one\cite{sauls_vortices_2009, daino_oddfrequency_2012}, having the topological charge $Q=2$\cite{garaud_properties_2015}.  In this section, we investigate the topological structure and the electronic properties of the doubly quantized coreless vortex state (skyrmion state) $(L_+, L_-, Q)=(0,2,2)$ and the vortex-skyrmion coexisting state $(L_+, L_-, Q)=(1,3,2)$.  We set parameters the same as in the previous section to facilitate the direct comparison of the results.  Note that we choose the $\Delta_-$-dominant states for convenience, so that the skyrmion corresponds to positive vorticity.  The $\Delta_+$-dominant counterpart with negative vorticity can be obtained equivalently by using Eq.~(\ref{Eq:BdG_TR2}).

We first present the topological structures of the state $(L_+, L_-, Q)=(0,2,2)$ in Fig.~\ref{skyr_tp}.  Fig.~\ref{skyr_tp}(a) shows $\Delta_{\pm}(r)$ and the supercurrent density profile $j_{\theta}(r)$.  Comparing to the results for the vortex free state $(L_+, L_-, Q)=(0,2,0)$ shown in Fig.~\ref{v0main}, a domain wall appears in $\Delta_{\pm}(r)$ at $r=12\xi_0$ separating outer $\Delta_-$ and inner $\Delta_+$ regions.  In addition, the winding numbers of $\Delta_{\pm}$ are $L_+=0$ and $L_-=2$, respectively.  There is therefore a $4\pi$-phase difference between $\Delta_{\pm}$ along the domain wall, which breaks the time reversal symmetry leading to the \textit{chiral domain wall}.  A supercurrent $j_{\theta}(r)$ is induced around the chiral domain wall, and changes sign at the domain wall - flowing clockwise inside the domain wall but anti-clockwise outside of it.\cite{fernandezbecerra_vortical_2016}  

The region inside the chiral domain wall is sometimes thought of as a vortex core.  However, this is not correct.  Different from the singular vortex which is a point-like topological defect, the coreless vortex is a loop-like topological defect.  Fig.~\ref{skyr_tp}(b) shows the results expressed using $\Delta_x$ and $\Delta_y$.  We found that $\Delta_x$ and $\Delta_y$ components of the OP contain two vortices each, thus having winding numbers $L_x=L_y=2$, so this state carries a total of 2 flux quanta.  The vortices are not at the sample center but on the chiral domain wall and align orthogonally in $\Delta_x$ compared to $\Delta_y$.  All four vortices are spatially separated and play the same role in this $(0,2,2)$ state, as seen from Fig.~\ref{skyr_tp}(b).  Therefore they are the one-component vortices (in $\Delta_x$-$\Delta_y$ space) and each of them carries half of the flux quantum, analogously to the half-quantum vortex.\cite{ivanov_nonabelian_2001}  Finally, the chiral domain wall is formed by an \textit{enclosed chain of all one-component vortices} and carries 2 flux quanta. The total OP is cylindrically symmetric, and it is suppressed (though not completely) on the chiral domain wall.  The relative phase $\theta_x-\theta_y$ alternates between 0 and $\pi$ along the domain wall, indicating that $\Delta_x$ and $\Delta_y$ are respectively in- and out of phase.  Note that the relative phase alternates exactly 4 times along the domain wall, where each node corresponds to the location of one-component vortices on the chiral domain wall.

Actually, the \textit{chiral domain wall} in $\Delta_{\pm}$ and the \textit{enclosed chain of one-component vortices} in $\Delta_x$ and $\Delta_y$ are two different but both relevant aspects of a \textit{skyrmionic topological defect} in the relative OP space.  This can be seen clearly from Fig.~\ref{skyr_tp}(c) where we map both $\Delta_{\pm}$ and $\Delta_{x,y}$ decompositions of the OP onto the pseudo-spin fields $\mathbf{n}$.     
As seen from the upper panel, where the results are obtained by using OP components $\Delta_{\pm}$, the field $\mathbf{n}$ rotates at the domain wall which separates the central region where $\mathbf{n}$ points up and the region outside of the domain wall where $\mathbf{n}$ points down.  In addition, the field $\mathbf{n}$ rotates along the domain wall by $4\pi$, resulting in the nontrivial topological charge density on the chiral domain wall [see Fig.~\ref{skyr_tp}(d)].  The net topological charge $Q=2$ indicates that the field $\mathbf{n}$ wraps twice on the surface of the sphere [see Fig.~\ref{skyrmodel}(b)].  
The lower panel of Fig.~\ref{skyr_tp}(c) shows the results obtained by using OP components $\Delta_{x}$ and $\Delta_{y}$.  The field $\mathbf{n}$ also rotates at the domain wall.  In this case, the domain wall separates the central region where $\mathbf{n}$ points in positive $y$-direction and the outside region where $\mathbf{n}$ points in negative $y$-direction.  $\mathbf{n}$ also rotates by $4\pi$ along the domain wall, leading to the net topological charge $Q=2$.  In fact, this pattern can be reached by rotating the previous $n$ field by an angle $90^{\circ}$ about the $y$-axis.  The topological charge density and the net topological charge are invariant under this operation.  As a result, one concludes that $(0,2,2)$ state is a skyrmionic topological defect with $Q=2$ in the relative OP space, and that such topological structures retain the skyrmionic character under the transformation between $(\Delta_+,\Delta_-)$ and $(\Delta_x,\Delta_y)$ representations.

Next we present the electronic properties of this skyrmionic topological defect in the $(0,2,2)$ state in Fig.~\ref{skyr_elek}.  Previous studies revealed low energy excitations at the domain wall.\cite{sauls_vortices_2009, daino_oddfrequency_2012}.  However, the complete picture of excitations and LDOS is still lacking.  Here, our self-consistent calculations provide the more details of the quasiparticle excitation spectra and LDOS, enabling their identification in e.g. scanning tunneling microscopy (STM). 

Fig.~\ref{skyr_elek}(a) shows the quasiparticle excitation spectrum $E_n(\mu_n)$ and the corresponding LDOS$(r,E)$ near the domain wall.  As seen from Fig.~\ref{skyr_elek}(a), there are three distinct branches of bound states.  These are the surface bound states (S), the domain-wall bound states (DW) and the Andreev bound states (A).  The surface bound states are the same as those found in the vortex free states $(0,2,0)$, which were shown in Fig.~\ref{v0main}.  The domain-wall bound states and the Andreev bound states are typical for the skyrmion, i.e. chiral domain wall.

The domain-wall bound states cross zero energy with the lowest energy level having a small gap of the order $\Delta_0^2/E_F$.\cite{daino_oddfrequency_2012, mizushima_vortex_2010}.  Thus, the zero-energy Majorana states do not appear.  However,  the domain-wall bound states cause two effects in LDOS: a zero-bias peak at the domain wall, and the peak splitting with increasing or decreasing the bias.  One of those peaks shifts towards the interior of the domain wall, while the other shifts outward.  This feature can be seen clearly in Fig.~\ref{skyr_elek}(d), where we display the profile of the LDOS$(r)$ for bias energies $E/\Delta_0=0.3$, $0$, and $-0.3$.  

The Andreev bound states are induced near the gap energies $E \approx |\Delta_0|$, leading to peaks in LDOS at the domain wall, as seen from Fig.~\ref{skyr_elek}(c).  They are essentially similar to the quantum rotor state which is induced by multiple Andreev reflections at the normal/superconducting interface.\cite{lin_quantum_2014}  In that case, due to the time-reversal symmetry, Andreev bound states appear near both $E = \pm |\Delta_0|$.  However, the chiral domain wall breaks the time-reversal symmetry so that the Andreev bound states near $E = - |\Delta_0|$ are suppressed.

In addition, we found that the domain-wall bound states are electron-dominant (with spectral weight $Z_n < 0.5$) when they cross the zero bias, while the Andreev bound states are hole-dominant (with spectral weight $Z_n > 0.5$), as seen from Fig.~\ref{skyr_elek}(a) where the color coding indicates the spectral weight $Z_n$.  This feature can be seen clearly in Fig.~\ref{skyr_elek}(b), where we displayed the spectral weight $Z_n$ for all three types of bound states.  The domain wall bound states and the Andreev bound states are different from the surface bound states whose spectral weight is $Z_n = 0.5$.  These two branches of bound states are also different from the singly-quantized vortex bound states of $(L_+, L_-, Q)= (1,3,0)$ and $(L_+, L_-, Q)= (-1,1,0)$ shown in Fig.~\ref{singlemain}, which are fully coupled Bogoliubov quasiparticles with spectral weight $Z_n = 0.5$.

Due to the electron-dominant domain-wall bound states and the hole-dominant Andreev bound states, the LDOS near the domain wall exhibits asymmetry for bias energy $E \leftrightarrow -E$, as visible in Fig.~\ref{skyr_elek}(c).  This feature can be seen clearly in Fig.~\ref{skyr_elek}(e), where we displayed the LDOS$(E)$ as a function of bias energy at several radial distances $r$.  When we scan the LDOS far away from the chiral domain wall, e.g. at $r/\xi_0=5$, the superconducting coherence peaks are well established at the gap energy $\Delta_0$ and there is no LDOS peak when $|E|<\Delta_0$.  When $r/\xi_0=11$ (near the domain wall at $r_{skyr}/\xi_0=12$),  there are four peaks inside the gap energy $|E|<\Delta_0$.  Two of them are induced by the domain wall bound states [labeled by solid and open triangles in Fig.~\ref{skyr_elek}(e)].  The other two are induced by the Andreev bound states [labeled by diamonds in Fig.~\ref{skyr_elek}(e)].  Due to the electron-dominant domain-wall bound states, the peaks labeled by solid triangle have a higher amplitude than the ones labeled by the open triangle, which results in the asymmetric profile in LDOS.  At larger $r$, the two peaks labeled by triangles move towards each other and merge at the domain wall where $r/\xi_0=12$.  Simultaneously, the Andreev peak in negative $E$ labeled by diamond is significant due to the hole-dominant Andreev bound states, leading to another asymmetric profile in the LDOS.  When $r$ is further increased, the peaks labeled by triangles continue shifting and finally merge into the coherence peaks at gap energy $|E|=\Delta_0$.

\begin{figure*}
	\centering
	\includegraphics[width=\textwidth]{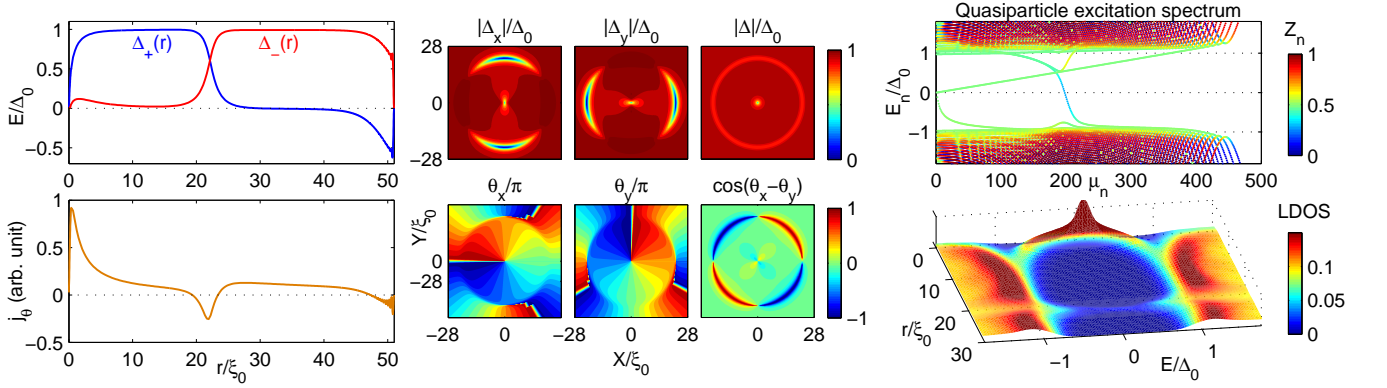}
	\caption{(Color online) The skyrmion-vortex coexisting state $(L_+, L_-, Q)=(1,3,2)$.  Plots on the left show profiles of $\Delta_{\pm}(r)$ and the azimuthal supercurrent density $j_{\theta}(r)$. Central plots show both amplitude and phase of OP components $\Delta_x(\mathbf{r})$ and $\Delta_y(\mathbf{r})$, their relative phase $\cos(\theta_x-\theta_y)$, and the total OP amplitude $|\Delta(\mathbf{r})|$.  Note that the winding numbers of $\Delta_x$ and $\Delta_y$ are $L_x=L_y=3$.  Plots on the right show the quasiparticle excitation spectrum $E_n$ as a function of the angular momentum $\mu$ (with color coding indicating the spectral weight $Z_n$), and the LDOS around the vortex core as a function of radial distance $r$ and bias energy $E$.}
	\label{skyrmionvortexmain}
\end{figure*}

Since the skyrmionic topological defect appears in the relative OP space, whereas the vortex appears in the OP space, a vortex can be added to the $(L_+, L_-, Q)=(0,2,2)$ state leading to the skyrmion-vortex coexisting state $(L_+, L_-, Q)=(1,3,2)$.  The results for such a topological ``hybrid'' are presented in Fig.~\ref{skyrmionvortexmain}(b).  Comparing to the skyrmion $(0,2,2)$ state, one sees the superposition of a singly quantized vortex and the chiral domain wall, with the vortex being located at center of the sample. The supercurrent $j_{\theta}(r)$ flows clockwise around the vortex core, gradually changing to anti-clockwise on the inner side of the domain wall, and flips the direction again to clockwise outside the domain wall.  $\Delta_x$ and $\Delta_y$ have winding numbers $L_x=L_y=1+2=3$ in this case, 1 for the central vortex, and 2 for the one-component vortices on the domain wall.  The chiral domain wall is larger than that of the skyrmion in the $(0,2,2)$ state, because of the repulsion between the vortex at the center and the one-component vortices on the domain wall.

The quasiparticle excitation spectrum $E_n(\mu_n)$ also shows the superposition of the vortex bound states and the chiral domain wall bound states.  Since the domain wall is now larger, the domain wall bound states and the Andreev bound states shift to larger $\mu_n$.  In addition, we find that the domain wall bound states become even more electron-dominant and the Andreev ones more hole-dominant, resulting in more pronounced electron-hole asymmetry in LDOS around the domain wall compared to the skyrmion $(0,2,2)$ state.  The LDOS of the coexisting skyrmion-vortex state exhibits distinctly strong zero-bias peak at the vortex core, and a significantly weaker one at the domain wall.

Finally, we mention that the skyrmion-anti-vortex coexisting state $(L_+, L_-, Q)=(-1,1,2)$ is unstable.  Due to the attractive interaction between the anti-vortex and the skyrmion, such state evolves into the parallel vortex state $(L_+, L_-, Q)=(1,3,0)$.

\section{Magnetic field and temperature dependence of the properties of the skyrmion}\label{sec:5}

The skyrmion is a chiral domain wall in $\Delta_{\pm}$ and an enclosed chain of one-component vortices in $\Delta_x,\Delta_y$ representation of the two component OP.  In either case, the skyrmion is a loop-like structure in OP space and it has very different properties from the vortex as a point-like defect.  For example, the size of the vortex depends solely on the superconducting coherence length $\xi$.  However, the size of the skyrmion depends also on the applied magnetic field because the chiral domain wall is expected to move under the influence of the magnetic field.  We therefore report in this section the magnetic field and temperature dependence of the size of the skyrmion in the $(L_+, L_-, Q)=(0,2,2)$ state, and the consequences of varied skyrmion size on the energy spectrum.

\begin{figure}
	\centering
	\includegraphics[width=\columnwidth]{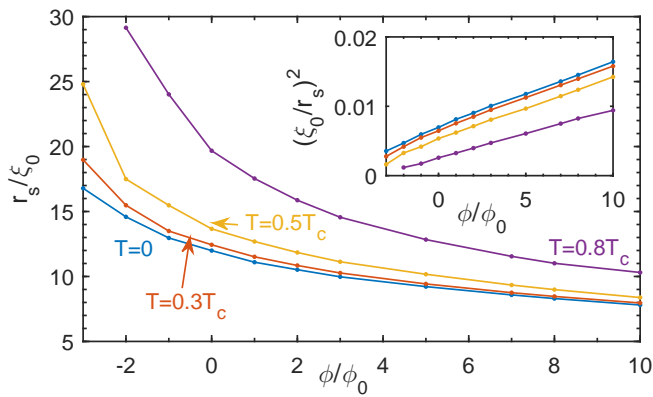}
	\caption{(Color online) The radius of the skyrmion $r_s$ as a function of the applied magnetic flux $\phi$ through the sample, at temperatures $T=0,~0.3,~0.5,$ and $0.8T_c$. The inset shows that the area of the skyrmion shrinks linearly with magnetic field, being maximal for negative magnetic field.}.
	\label{rsfit}
\end{figure}

Fig.~\ref{rsfit} shows the radius $r_s$ of the $\Delta_-$-dominated $(0,2,2)$ skyrmion, as a function of the magnetic flux $\phi$ through the sample, at temperatures $T=0,~0.3,~0.5,$ and $0.8T_c$.  The $\phi = H_0S$ where $H_0$ is the magnetic field strength and $S=\pi R^2$ the area of the sample.  We find that the skyrmion expands with increasing temperature $T$, but shrinks with increasing applied magnetic field.  The skyrmion consists of the one-component vortices, with size related to the coherence length $\xi$.  Since $\xi$ increases with temperature, so does the vortex-vortex interaction, and the size of the skyrmion can duly increase.  However, it is crucial here that the skyrmion is a chiral domain wall, balanced by the clockwise supercurrnt $j_{\theta}$ in the interior and the anti-clockwise at the exterior of the domain wall.  With increasing applied magnetic field, the anti-clockwise part of $j_{\theta}$ is enhanced and the clockwise part is weakened, shrinking the domain wall to smaller equilibrium radius $r_s$. Inversely, the skyrmion expands with $\phi$ decreasing.  Interestingly, the skyrmion survives even at negative magnetic field, i.e. for $\phi<0$, likely due to the finite energy needed to break the domain wall so that vortices can leave the sample. As a consequence, at negative fields skyrmion continues to expand to surprisingly large sizes.  The inset in Fig.~\ref{rsfit} shows that actually the square of $r_s(\phi)$ depends linearly on $\phi$, i.e. $\propto 1/\phi^2$, so that magnetic flux inside the skyrmion is roughly constant. This is a very important finding, indicating that existing skyrmions in a given sample can be made larger, hence easier to detect in experiment, if the polarity of the applied magnetic field is reversed. Furthermore, the stability at reversed field clearly distinguishes skyrmions from vortices, since there is nothing preventing individual vortices from leaving the sample (apart from the ever-present disorder) if the polarity of the field is changed. Last but not least, our findings indicate that skyrmions are in general \textit{an order of magnitude larger} than the conventional vortices.

\begin{figure}
	\centering
	\includegraphics[width=\columnwidth]{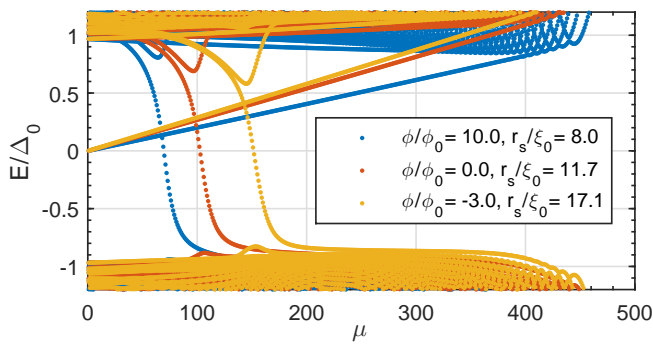}
	\caption{(Color online) Quasiparticle excitation spectrum $E_n$ of the skyrmion studied in Fig. \ref{rsfit}, as a function of angular momentum $\mu_n$, at zero temperature and for applied magnetic flux through the sample $\phi/\phi_0=-3$, 0, and $10$.}
	\label{Ekrs}
\end{figure}
\begin{figure*}
	\centering
	\includegraphics[width=1\textwidth]{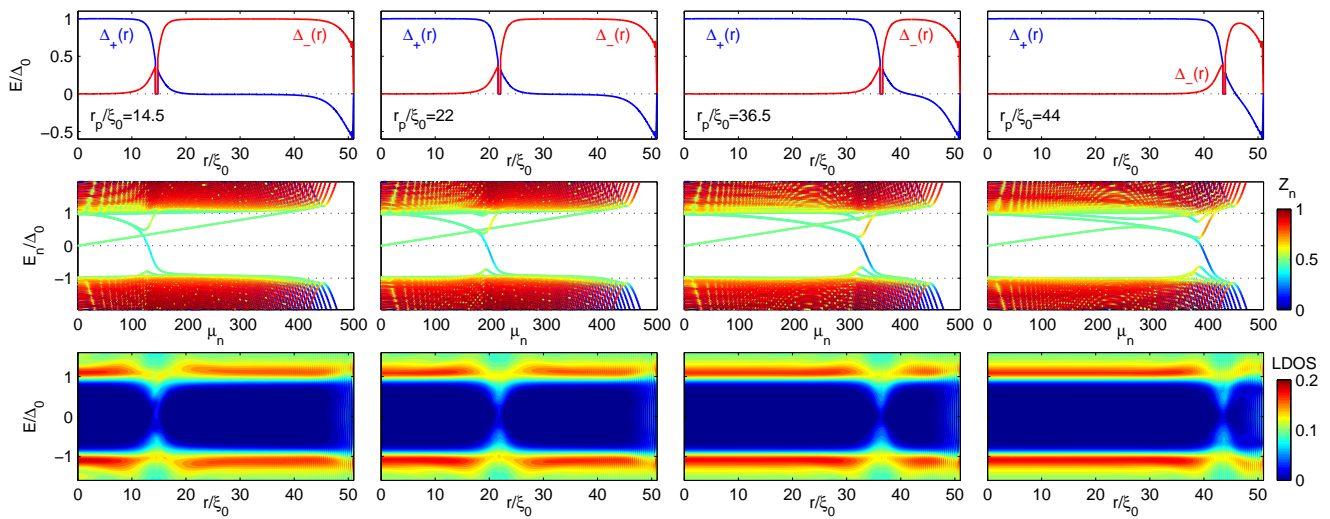}
	\caption{(Color online) Skyrmion $(L_+, L_-, Q)=(0,2,2)$ state trapped by a normal-metal ring.  The radius of the pinning rings increases as $r_p/\xi_0=14.5, 22, 36.5, 44$ from left to right panels, respectively.  The top row of plots shows the OP profiles.  The central row shows the corresponding quasiparticle excitation spectrum as a function of angular momentum $\mu_n$, and the bottom row shows the LDOS as a function of radial distance $r$.}
	\label{pinning}
\end{figure*}

The electronic structure is of course affected by the change in the size of the skyrmion.  Fig.~\ref{Ekrs} shows the quasiparticle excitation spectrum $E_n(\mu_n)$ of the skyrmion at zero temperature, for magnetic flux through the sample $\phi/\phi_0=10$, 0, and $-3$, for which $r_s/\xi_0=8$, $11.7$ and $17.1$, respectively.  The domain-wall bound states move to large angular momentum $\mu$ when $r_s$ increases, which is expected since the bound states are confined to the domain wall.  In addition, the cusped energy lines of the Andreev bound states become more significant around $E=|\Delta_0|$.  The continuous spectrum above the gap energy $|E|>|\Delta_0|$ tilts as a function of $\mu_n$ because of the supercurrent induced by the applied magnetic field favoring one chirality over the other.

\section{Pinning the skyrmion}\label{sec:6}

Vortex matter in superconductivity is known to be pinned where the OP is suppressed, which can have technological relevance for e.g. increasing the maximal current a superconductor can sustain without the onset of vortex motion and related onset of resistance and heating.  The skyrmion matter is a chain of enclosed one-component vortices according to the OP representation using $\Delta_x$ and $\Delta_y$, implying that skyrmions can be pinned in an analogy to vortices.  If so, then the size and the position of the skyrmion could be controlled artificially, which may be beneficial for the observation of skyrmions and for further fluxonic manipulations.  In this section, we therefore consider the possibility to pin the skyrmion by an embedded normal-metal ring in the superconductor, where the superconducting coupling constant $g$ is suppressed to zero, leading to $|\Delta|=0$ inside the ring.  The median radius of the ring is labeled $r_p$, and the width of the ring is $0.5\xi_0$. Such narrow rings do not break the phase coherence between the superconductivity inside and outside of the ring.  We investigate the OP profile, energy spectrum and LDOS when the skyrmion is pinned by such a normal-metal ring.  The calculations are performed self-consistently for $T=0.1T_c$ and in absence of the magnetic field, since we do not want the competing effects to shadow the conclusions.

Fig.~\ref{pinning} presents the OP profiles (top row), quasiparticle excitation spectrum (central row) and LDOS (bottom row) for the radii of the normal-metal ring $r_p/\xi_0=14.5, 22, 36.5, 44$ (from left to right respectively). A seen in the OP profiles in Fig.~\ref{pinning}, the chiral domain walls are trapped in the normal-metal ring in every considered case.  With increasing radius of the ring $r_p$, the skyrmion correspondingly expands.  As a result, the domain wall bound states shift to larger angular momentum $\mu_n$ in the energy spectrum, and the zero-bias peak in LDOS shifts as well.  Note that the domain wall bound states become increasingly hole-dominant with the expansion of the skyrmion.  At the same time, the Andreev bound states around $E=|\Delta_0|$ become more significant and increasingly electron-dominant.

The surface bound states are not affected by our exercise until the skyrmion gets close to the sample surface. As seen from the panels for $r_p/\xi_0=44$, the OP profiles at the surface are strongly affected by the domain wall.  The supercurrents induced by the domain wall and ones running near the surface combine, causing interactions between the domain wall bound states and the surface bound states.  As seen from the energy spectrum $E_n(\mu_n)$, these two branches of bound states avoid crossing each other.  Finally, we note that the quasiparticles interference above the gap energy $|E|>\Delta_0$ is enhanced with the $r_p$ increasing.  The quasiparticles interference effect is known to result in additional BCS-like energy gaps and more Bogoliubov quasiparticle states with $Z_n=0.5$ above the gap energy $\Delta_0$.\cite{zhang_tomasch_2015}  Here, it is induced by the inhomogeneous OP profile stemming from the normal-metal ring, the skyrmion and the surface.

\section{Summary}\label{sec:7}

In summary, we have studied the topological and electronic properties of characteristic vortical and skyrmionic states in chiral $p$-wave superconductors, by solving Bogoliubov-de Gennes equations self-consistently. We have presented the distribution of the two-component order parameter, the supercurrent, quasiparticle excitation spectra, and LDOS, for each of the typical states.  We pointed out that the chiral order parameter representation using components $\Delta_{\pm}=p_x\pm ip_y$ is ideal to study the properties of chiral domain walls in the given state, while the $p_x$- and $p_y$-components of the order parameter conveniently reveal the properties of vortices.

While conventional vortices are rather well understood in the literature (as point-like topological defects, with core in the order parameter, supercurrent flow around it, and the vortex bound states and LDOS peaks at the core), the topological defects comprising one-component vortices, and/or chiral domain walls as well as their interaction with conventional vortices, are an entirely new topic.  Moreover, a chain of one-component vortices (half the vorticity of a complete vortex, analogous to half-quantum vortices of spin-triplet superconductors\cite{chung_stability_2007}) on a chiral domain wall can be characterized as a skyrmion, and can be seen in the total order parameter as loop-like topological defect without a fully developed core.  Such defects carry multiple flux quanta, but are entirely different from ``giant'' vortices in $s$-wave superconductors.\cite{schweigert_vortex_1998, virtanen_multiquantum_1999, cren_vortex_2011} Such skyrmion exhibits a chiral domain wall in $\Delta_{\pm}$, whereas a vortex does not. Unlike vortices, they are characterized not only by the angular momentum, but also by the topological charge in the relative order parameter space, where both the relative amplitude and relative phase between the two components of the order parameter play a role.  A skyrmion traps bound states at the chiral domain wall, leading to zero-bias LDOS peaks at the domain wall.  In addition, the LDOS exhibits electron-hole asymmetry, which is different from the electron-hole symmetric LDOS of usual multi-quanta vortex states.  We also show the possibility to have a topological defect with a vortex inside a skyrmion, with superimposed features of both topological constituents.  

Our analysis in varied magnetic field and temperature shows that the size of the skyrmion can be strongly tuned, being increased by increasing temperature and by decreasing applied magnetic field. The size of the skyrmion is typically an order of magnitude larger than a vortex. Furthermore, contrary to conventional vortices, a skyrmion survives changing the polarity of the applied magnetic field, due to the finite energy cost of breaking the chiral domain wall so that vortices within the skyrmion can leave the sample. As a consequence, the skyrmion can significantly increase in size at negative magnetic field, since the decreasing energy of currents flowing inside the skyrmion compensates the increasing energy of the longer chiral domain wall. Finally, we have shown that even in the absence of the magnetic field the size of the skyrmion can be manipulated by pinning on a normal-metal ring of prescribed size. Considering that due to recent experimental achievements in e.g. superconductor-ferroelectric hybrids one can draw practically at will the normal-metal paths inside the superconductor,\cite{villegas_bistability_2007, visani_hysteretic_2011} this opens up a broad playground for novel phenomena in fluxonics. We expect that our findings related to stability of skyrmionic topological defects in superconductors, manipulation of their size, and their distinct signatures in for example LDOS, will enable their experimental identification in scanning tunneling microscopy and spectroscopy, which can be further used to prove particular pairing symmetry in the superconductor of interest.

\section*{Acknowledgments}
This work was supported by the Research Foundation-Flanders (FWO-Vlaanderen).

\end{document}